\documentclass[aps,prd,amsfonts,amsmath,nofootinbib,showpacs,preprint,tightenlines,longbibliography]{revtex4-1}

\def\be{\begin{equation}}
\def\ee{\end{equation}}
\def\bea{\begin{eqnarray}}
\def\eea{\end{eqnarray}}
\def\bml{\begin{subequations}}

\def\elea{\end{eqnarray}\end{subequations}}
\def\gbb{\Gamma^{\text{BB}}}
\def\mbb{M_{\text{BB}}}
\def\mpl{M_{\text{Pl}}}
\def\mbh{M^{\text{BH}}_{\text{min}}}
\def\vnuc{V^{\text{nuc}}}
\def\mnuc{M^{\text{nuc}}}

\begin{document}

\title{Black holes and up-tunneling suppress Boltzmann brains}

\author{Ken D. Olum}
\email{kdo@cosmos.phy.tufts.edu}

\author{Param Upadhyay}
\email{pupadh01@tufts.edu}

\author{Alexander Vilenkin}
\email{vilenkin@cosmos.phy.tufts.edu}

\affiliation{Institute of Cosmology, Department of Physics and
  Astronomy, Tufts University, Medford, MA 02155, USA} 

\begin{abstract}
Eternally inflating universes lead to an infinite number of Boltzmann
brains but also an infinite number of ordinary observers.  If we use
the scale factor measure to regularize these infinities, the ordinary
observers dominate the Boltzmann brains if the vacuum decay rate of
each vacuum is larger than its Boltzmann brain nucleation rate.  Here
we point out that nucleation of small black holes should be counted in
the vacuum decay rate, and this rate is always larger than the
Boltzmann brain rate, if the minimum Boltzmann brain mass is more than
the Planck mass.  We also discuss nucleation of small, rapidly
inflating regions, which may also have a higher rate than Boltzmann
brains.  This process also affects the distribution of the different
vacua in eternal inflation.

\end{abstract}

\maketitle

\section{Introduction}

If the observed dark energy is in fact a cosmological constant, our
universe will expand forever and will soon approach de Sitter space.
There will be an infinite volume in which many types of objects may
nucleate.  In particular there will be an infinite number of Boltzmann
brains \cite{Albrecht:2004ke}, human brains (or perhaps computers that
simulate brains) complete with our exact memories and thoughts, that
appear randomly as quantum fluctuations.  Human beings (and their
artifacts) can arise in the ordinary way for only a certain period of
time after the Big Bang, when there are still stars and other
necessities of life, but Boltzmann brains can arise at any time in the
future.  So one might conclude that the Boltzmann brains infinitely
outnumber ordinary humans, and thus that we are Boltzmann brains, a
nonsensical conclusion \cite{Carroll:2017gkl} because our observations
on which we base this conclusion would have no connection to the
actual universe in which we live.

However, in any scenario such as the above, it is also possible for
new inflating regions to nucleate, leading to eternal inflation.  In
that case there will be an infinite number of ordinary observers in
addition to the infinite number of Boltzmann brains.  In order to know
what to expect in such situations we need a measure: a procedure to
regulate the infinities and produce a sensible probability
distribution.  Any measure faces a number of difficulties
\cite{Bousso:2010yn,Guth:2011ie,Olum:2012bn} and we do not have any
principle to tell us which measure is correct.  An obvious selection
criterion is that the measure should not make predictions that are in
conflict with observation.  This removes most of the measures that
have been suggested so far.  The proper time measure suffers from the
"youngness paradox", predicting that the CMB temperature should be
much higher than observed \cite{Tegmark:2004qd}; the causal patch
measure predicts that the cosmological constant should be negative
with an overwhelming probability \cite{Bousso:2010im}; the pocket
based measure suffers from a ``$Q$-catastrophe'', predicting either
extremely small or large values of the density fluctuation amplitude
$Q$ \cite{Feldstein:2005bm,Garriga:2005ee}.  A measure that fares
reasonably well is the scale factor cutoff measure
\cite{Linde:1993xx,DeSimone:2008bq,Bousso:2008hz}.  Other measures
that have not been ruled out by observations (such as the lightcone
time cutoff, apparent horizon cutoff and 4-volume cutoff measures)
make predictions very similar to the scale factor cutoff.  (For more
details and references see, e.g., Ref.~\cite{Freivogel:2011eg}.)

In the present paper we shall adopt the scale factor cutoff measure,
which we will discuss in more detail below.  In this measure, the
ratio of Boltzmann brains to ordinary observers in a given vacuum is
roughly given by the ratio of the Boltzmann brain nucleation rate
$\gbb_i$ to the total decay rate of that vacuum $\Gamma_i$.  Here
$\gbb_i$ is the rate at which Boltzmann brains form per unit
(physical) volume of vacuum $i$, and $\Gamma_i$ is proportional to
the total rate at which volume flows out of vacuum i (a precise
definition of $\Gamma_i$ will be given below).

We point out here that there are two processes that are not always
considered that influence the vacuum decay rate.  The first is the
nucleation of small black holes.  This process removes volume from the
vacuum, and so contributes to $\Gamma_i$.  The rate is largest for the
smallest black holes.  As we will discuss below, it is always larger 
than the Boltzmann brain nucleation rate,
if the minimum Boltzmann brain mass is larger than the Planck mass,\footnote{A similar argument was made in
  Ref.~\cite{Garriga:2012bc} in the context of the "watcher measure".
  This measure makes the assumption that the big crunch singularities
  in AdS bubbles lead to bounces, where contraction is followed by
  expansion, so that geodesics can be continued through the crunch
  regions. We do not adopt this assumption in the present paper.}
so the Boltzmann brain problem is solved in that case.

The other process is the nucleation of small regions of higher-energy
inflating false vacuum.  In the usual Lee-Weinberg \cite{Lee:1987qc}
process, a region larger than the Hubble distance in the old vacuum
tunnels to the new vacuum.  But here we are considering a localized
fluctuation that yields a region of the new vacuum large enough to
inflate but much smaller than the old Hubble distance
\cite{Brown:2011ry,Blanco-Pillado:2019xny}.  The higher the energy of
the new vacuum, the smaller the region of it that is necessary for
inflation.  Thus this process (unlike Lee-Weinberg tunneling) is least
suppressed when the daughter vacuum energy is the highest.  The most
likely process is to produce the highest energy inflating vacuum.  If
this is at the Planck scale, suppression is similar to that of
Planck-scale black hole production.  Otherwise it is more suppressed
than that.

Nucleation of small high-energy regions is not discussed in most
treatments of the multiverse physics.  We shall comment on the reason
for that below and explain why we believe it should be included.  This
process upends the conventional wisdom that low-energy vacua are most
likely to tunnel to other low-energy vacua.  Up-tunneling is still 
suppressed when the parent vacuum energy is small, but now the most
likely daughters are the ones with the highest energy.  To compute the
probabilities in the scale factor measure, we construct a transition
matrix between vacua and find its eigenvector whose eigenvalue is
least negative.  This is usually made up almost entirely by a single
``dominant vacuum'' \cite{SchwartzPerlov:2006hi,Olum:2007yk} whose
total decay rate is the least.  The measures of other vacua depend on
tunneling processes leading to them from the dominant vacuum.  When we
take into account production of small high-energy inflating regions,
we can still find the dominant vacuum, but the details do not matter.
The likeliest transition out of the dominant vacuum is to jump
directly to the highest energy possible.  At very high energies,
transitions are little suppressed, so all vacua are quickly populated.  The
chance of any specific low-energy (and in particular anthropically
allowed) vacuum depends now on how it may be reached by a sequence of
transitions from high energies, with little effect from the details of
the dominant vacuum.

The rest of this paper is organized as follows.  In the next section
we review the scale factor measure and the resulting distribution
of the different vacua, and discuss the effects of nucleating
black holes and small high-energy regions.  In
Sec.~\ref{sec:nucleation} we discuss the nucleation rates of black
holes, Boltzmann brains, and regions of different vacuum.  We discuss
the effects of these processes on the Boltzmann brain problem in
Sec.~\ref{sec:BB} and on the distribution of the different vacua in
Sec.~\ref{sec:vacuum}.  We conclude in Sec.~\ref{sec:conclusions}.

\section{The scale factor cutoff}\label{sec:scalefactor}

The scale factor measure was introduced by Linde and collaborators
(e.g., \cite{Linde:1993xx}) and was worked out in detail in
Refs. \cite{DeSimone:2008bq,Bousso:2008hz,DeSimone:2008if}.  It is
based on constructing a scale factor time that represents (the
logarithm of) the total expansion that each point in spacetime has
experienced.  To make it well defined, we must start with some initial
spacelike hypersurface $\Sigma$ and follow a congruence of geodesics
orthogonal to $\Sigma$.  The scale factor time is then given by
\be
\eta = \int_0^t\frac{\theta}{3} dt'\,,
\ee
where $t$ is proper time, and the expansion $\theta = {u^\mu}_{;\mu}$,
with $u^\mu = dx^\mu/dt$ the tangent vector to the geodesics.
In a homogeneous region of the universe, the scale factor $a$ is just
$\exp\eta$.

To use this as a measure, we consider all events that take place
before some cutoff time $\eta_c$.  There are a finite number of these,
so assigning probabilities is straightforward.  Then we take the
limit of the probabilities as $\eta_c$ grows without bound.

Unfortunately, when structures form, the local universe contracts
instead of expanding, so $\eta$ is not monotonic; we must make some
provision for this case \cite{Bousso:2008hz}.  One plan would be to
use a modified scale factor time $\tilde\eta$, where $\tilde\eta(x)$
is given by maximizing $\eta$ over all points in the causal past of
$x$.  Thus $\tilde\eta$ cannot decrease, and we avoid the possibility
that an event allowed by the cutoff is in the future of a point
excluded by the cutoff.\footnote{Such a situation would lead to an inverse
Guth-Vanchurin \cite{Guth:2011ie} paradox where an observer may wake
up without ever having gone to sleep.}  A number of other possibilities have
been suggested \cite{Bousso:2008hz,DeSimone:2008if}.

It will not be important to our analysis here exactly how this issue
is resolved, but for definiteness we will use the above ``maximum
$\eta$'' prescription.  At any given scale factor, almost all the
volume is in regions that are expanding.  In such a region, the
distance between two geodesics of the congruence is just their
distance on the initial surface times the expansion of the scale
factor.  If we select an evenly spaced, very large but finite set of
representative geodesics on initial surface, all of these geodesics
will represent equal volumes on the cutoff surface.  Thus the fraction
of volume in each type of region is just the fraction of the initial
geodesics that are there.

We will be interested in the number of Boltzmann brains and
ordinary observers that appear in different vacua.  Let us start by
defining $f_i(\eta)$ to be the fraction of comoving volume in vacuum
$i$ at time $\eta$.  In expanding regions, the expansion factor $a$ is
the same everywhere on the constant-$\eta$ surface, so $f_i$ gives the
fraction of physical volume.  In contracting regions, we must make
some adjustment, as described above.  But such regions will not
matter, as we discuss later.

The $f_i$ obey the rate equation \cite{SchwartzPerlov:2006hi},
\be\label{eqn:rate}
\frac{df_j}{d\eta} = \sum_i\left (-\kappa_{ij} f_j+\kappa_{ji} f_i\right)\,,
\ee
where $\kappa_{ij}$ is the fraction of volume currently in vacuum $j$
that transitions into vacuum $i$ per unit scale factor time, or
equivalently the chance per unit scale factor time for an
observer in vacuum $j$ to transition to vacuum $i$.

We can express $\kappa_{ij}$ in terms of $\Gamma_{ij}$, the rate of
tunneling events that produce vacuum $i$ per unit physical spacetime
volume of vacuum $j$.  In general,
\be
\kappa_{ij} = \frac{V_{ij}}{H_j} \Gamma_{ij}\,,
\label{kappa}
\ee
where $V_{ij}$ is the volume of space at a given time where a given
tunneling event would lead to a given observer transitioning to the
new vacuum.  The expansion rate $H_j$ of vacuum $j$ in the denominator
is the conversion between scale factor time and physical time.

In the Coleman-De Luccia process, a small region of lower-energy
vacuum appears by tunneling and then expands to the horizon size.  In
the Lee-Weinberg process, a super-horizon region of a higher-energy
vacuum appears by tunneling and then contracts in comoving size so the
final comoving volume is just the comoving horizon at the time of
nucleation.  In either case, $V_{ij} = (4\pi/3) H_j^{-3}$.

Here we will discuss two more processes.  The first is the nucleation
of black holes.  A certain set of geodesics will fall into the black
hole, hit the singularity, and be removed from the congruence.  They
will reach a maximum $\eta$ before they start to converge near the
black hole; for larger $\eta$ they will not be counted in the scale
factor measure.  They thus represent a flow of volume fraction out of
the vacuum in which the black holes nucleate.  In that sense the
process is similar to the creation of anti-de Sitter vacua that then
collapse.  We will describe black hole nucleation by a transition rate
$\kappa_{0j}$ for each vacuum $j$, and include it in
Eq.~(\ref{eqn:rate}) by including $i=0$ in the sum.

If a black hole of mass $M$ lives for a time long compared to the
Hubble time, it will capture all geodesics within radius \cite{Garriga:2012bc}
\be
r_c \sim \left(\frac{GM}{H^2}\right)^{1/3}\,,
\ee
which is the radius at which the attraction of the black hole
gravity is balanced by the repulsive force due to the cosmological
constant.  The volume of geodesics absorbed is thus
\be
V_c \sim \frac{GM}{H^2}\,.
\ee

If the black hole is short-lived compared to the Hubble
time, we can neglect the cosmological constant.  A particle starting
from rest at radius $r$ will fall into the black hole on a time scale
\be
t \sim r^{3/2} (GM)^{-1/2}\,.
\ee
We want $t < t_e$, where the evaporation time is\footnote{We note
  that magnetically charged black holes may be much more stable.  They
  can lose their magnetic charge only by emission of magnetic
  monopoles, which typically have large masses, so their emission may
  be strongly suppressed.  The black hole may even be absolutely
  stable if monopole solutions of corresponding magnetic charge do not
  exist.}  $t_e \sim G^2 M^3$.  From this we find that the capture
radius and volume are
\bea\label{eqn:rcsmall}
r_c  &\sim&  G^{5/3} M^{7/3}  \sim  GM \left(\frac{M}{\mpl}\right)^{4/3}\\
V_c &\sim& G^5 M^7 \sim (GM)^3 \left(\frac{M}{\mpl}\right)^4\,.
\eea

We will also consider the formation of regions of higher cosmological
constant $\Lambda_i$ that are smaller than the horizon of the parent
vacuum $j$, but larger than their own horizon.  Such a region will
inflate inside, but the outside will collapse into a black hole.  As
we mentioned in the Introduction, this nucleation process is often
omitted in studies of multiverse dynamics.  The main reason is that it
does not fit into the standard Coleman-De Luccia formalism, where
tunneling transitions are described by instantons.  There are no known
instantons corresponding to nucleation of small high-energy inflating
regions.  However, quantum transitions allowed by the conservation
laws should occur with some nonzero probability.  The state of a
quantum field in de Sitter space is similar to a thermal state, and
one expects that fluctuations of the scalar field $\phi$ and/or its
velocity ${\dot\phi}$ will occur in localized regions of space.  If
the fluctuation is large enough, the field may acquire enough energy
to fly over a potential barrier into a high-energy vacuum.  And if the
fluctuation extends over a super-horizon region in the new vacuum, it
will produce an inflating baby universe
\cite{Brown:2011ry,Blanco-Pillado:2019xny}.

Geometrically it is clear that a rapidly inflating daughter region
will be connected by a wormhole to the slowly inflating parent
universe.  The wormhole will close up in about one light crossing time
and both of its mouths will be seen as black holes.\footnote{This
  process is similar to that described in Ref.~\cite{Garriga:2015fdk}.
  It is also related to the process of Ref.~\cite{Farhi:1989yr}, but
  in that paper the authors propose deliberately constructing a region
  of high-energy vacuum that is not large enough to inflate and hoping
  that it tunnels to the inflating state, while here we propose
  creating the region as a fluctuation in de Sitter space.}.  After
the black hole evaporates, the new inflating region is disconnected
from the original universe, but there is no problem in applying the
scale factor measure to the resulting set of disconnected universes.

If we could ignore gravitational effects, it would be easy to compute
the energy necessary to create such a region.  Let $U_i$ be the energy
density of the daughter vacuum.  The expansion rate is thus $H_i \sim
\sqrt{G U_i} = \sqrt{U_i}/\mpl$.  The minimal volume to inflate would
be a sphere of radius $H_i^{-1}$, which thus contains volume
\be\label{eqn:vnuc}
\vnuc_i \sim \mpl^3/U_i^{3/2}
\ee
and mass
\be\label{eqn:mnuc}
\mnuc_i \sim \mpl^3/U_i^{1/2}\,.
\ee
This is modified by gravitation, but we will assume here that the
effect is only to change the numerical factors that we did not compute
and so Eqs.~(\ref{eqn:vnuc}) and (\ref{eqn:mnuc}) give the correct
order of magnitude.

The new inflating volume must be surrounded by a bubble wall that
interpolates between the two vacua.  This is same wall as in
Lee-Weinberg and the inside-out version of the Coleman-De Luccia
bubble wall.  Suppose it is possible, as one normally expects, for a
small bubble of vacuum $j$ to form inside a Hubble volume of vacuum
$i$.  That means that the energy of this wall around a sphere of
radius smaller than $1/H_i$ is less than the energy of the displaced
volume of vacuum $i$.  In the present case, we have a larger sphere,
of radius $1/H_i$, which increases the ratio of volume to surface
energy.  So the wall energy will be much less than $\mnuc_i$, which is
the energy of the sphere of vacuum $i$ of radius $1/H_i$, and there is
no important correction to $\mnuc_i$, from the wall.

Upon nucleation we expect that the geodesics inside volume $\vnuc_i$
will travel into the new inflating region.\footnote{A geodesic
  congruence is not well defined when the spacetime undergoes a
  discontinuous change, as in quantum tunneling.  But it should be
  possible to estimate, by order of magnitude, what fraction of the
  initial comoving volume goes into each vacuum.  This is typically
  all one needs in any anthropic analysis.}  Shortly after that, the
nucleated region will collapse into a black hole, and more geodesics
will later fall into the black hole and end at the singularity,
according to Eq.~(\ref{eqn:rcsmall}).  So this process gives
\be
\kappa_{ij} \sim \frac{\vnuc_i}{H_j} \Gamma_{ij}
\sim \Gamma_{ij} H_j^{-1} H_i^{-3}
\label{Gamma1}
\ee 
and in addition contributes
\be
\sim \Gamma_{ij} H_j^{-1} \mpl^3 H_i^{-6}
\label{Gamma2}
\ee 
to $\kappa_{0j}$.

Calculation of the relative abundance of Boltzmann brains and ordinary
observers involves comparisons of extremely small numbers, like
tunneling transition rates $\kappa_{ij}$ and Boltzmann brain
nucleation rates.  The tunneling actions are typically large, so these
rates are double exponentially suppressed.  The pre-exponential
factors have therefore little effect, even though they can be very
small or large.  For this reason the factors multiplying $\Gamma_{ij}$
in Eqs.~(\ref{kappa}), (\ref{Gamma1}) and (\ref{Gamma2}) can be
ignored, and we will omit them from now on.

\section{Nucleation rates}\label{sec:nucleation}

In this section, we review the usual nucleation rates for the
Coleman-De Luccia and Lee-Weinberg cases and discuss nucleation of
black holes, small inflating regions, and Boltzmann brains.

\subsection{Coleman-De Luccia and Lee-Weinberg nucleation}

A metastable vacuum $j$ may decay to a lower energy vacuum $i$ through
bubble nucleation.  If we ignore the effects of gravitation, we have
the situation discussed by Coleman \cite{Coleman:1977py}.  It proceeds
by forming a bubble whose total energy is zero because the decreased
energy of the vacuum inside compensates for the energy in the bubble
wall.  Including gravitation \cite{Coleman:1980aw} leads to
corrections, but these are small if the bubble size is small compared
to the Hubble distance in both parent and daughter vacua.  After
formation, the bubble will expand rapidly because the force on the
wall due to the difference in vacuum energies is larger than the
effect of surface tension.  Disregarding the pre-exponential factor,
the bubble nucleation rate is given by
\be
\Gamma_{ij}\sim e^{-I-S_j}\,,
\ee
where $I<0$ is the instanton action and $S_j=\pi/H_j^2$ is the Gibbons-Hawking entropy of the parent vacuum $j$.

Lee and Weinberg \cite{Lee:1987qc} have argued that the same instanton should describe the inverse transition from $i$ to $j$, where the daughter vacuum has a higher energy than the parent vacuum.  The corresponding transition rate is
\be
\Gamma_{ji}\sim e^{-I-S_i}\,,
\ee
It follows that the upward and downward transition rates are related by
\be
\Gamma_{ji}/\Gamma_{ij}\sim e^{S_j-S_i}\,.
\label{balance}
\ee
If the two vacuum energies are significantly different, the upward transition rate is very strongly suppressed.
Eq.~(\ref{balance}) can be interpreted as an expression of detailed balance between vacuum transitions in the multiverse.  It fits well with the widely accepted picture of quantum de Sitter space as a thermal state \cite{Dyson:2002pf}.

Analytic continuation of the instanton to the Lorentzian regime indicates that in the case of upward tunneling the initial size of the bubble is larger than the parent vacuum horizon $H_i^{-1}$.
The high-energy bubble is pushed inward because the
vacuum energy density outside is smaller than the density inside, and
thus the inside pressure is more negative.  But since the bubble is
outside the Hubble distance it is carried outward by the Hubble
expansion, even though locally it accelerates inward.
        
\subsection{Black hole nucleation}

In general we expect an arbitrary object of mass $M$ that is much
smaller than the Hubble distance to appear in de Sitter space at a
rate proportional to
\be\label{eqn:thermal}
\exp(-2\pi M/H) = \exp(-M/T)\,.
\ee
The latter expression gives the likelihood of finding such an object
in a thermal bath in the Gibbons-Hawking temperature $T =
H/(2\pi)$.\footnote{More precisely, the nucleation rate is
  proportional to $\exp(-F/T)=\exp(-M/T+S)$, where $F$ is the free
  energy and $S$ is the entropy of the nucleating object.   This
    takes account of the possibility of nucleating the object in various
    microstates.  The correction, however, is small in cases of
  interest to us here.}  The former expression has been found by
instanton calculations, for example see Ref.~\cite{Basu:1991ig} for
the nucleation of monopoles and Ref.~\cite{Bousso:1998na} for the
nucleation of black holes.  The calculation of black hole nucleation
rate in Ref.~\cite{Bousso:1998na} is somewhat controversial, since it is
based on an instanton with a conical singularity.  Exclusion of such
instantons leads to the conclusion that only maximal black holes of
horizon radius equal to the cosmological horizon can nucleate in de
Sitter space \cite{Ginsparg:1982rs}.  However, regular instantons do
exist for nucleation of electrically or magnetically charged black
holes of sub-maximal mass \cite{Mellor:1989wc}.  In the limit of small
mass the corresponding nucleation rate is given by
Eq.~(\ref{eqn:thermal}).  We note also that Eq.~(\ref{eqn:thermal})
would give the rate to nucleate a distribution of dust that would
collapse into a black hole.

\subsection{Small inflating regions}

As discussed above, it is possible to nucleate a much smaller bubble
of higher energy vacuum $i$.  As seen from the outside, the force
on the bubble wall will cause it to shrink, leading the bubble to
collapse into a black hole.  However, if the bubble volume is larger
than $\vnuc_i$, it will inflate on the inside, leading to a
new inflating region of vacuum $i$.

What is the rate at which such regions are produced?  The simple
conjecture is that it is proportional to
\be
e^{-\mnuc_i/T} \sim e^{-\mpl^3/(T\sqrt{U_i})} \sim e^{-\mpl^2/(H_i H_j)}\,,
\label{flyover}
\ee
as we would expect for any object of mass $\mnuc_i$.  However, there
are some caveats.  New small inflating regions cannot be produced by
any classical process, because their production violates the null
energy condition \cite{Farhi:1986ty}.  Thus a classical thermal state
would not produce regions such as these, perhaps casting some doubt on
the use of a thermal expression above.  This is a fundamentally
quantum process, so perhaps it can be described by an instanton, but
such an instanton is not known.  A similar situation was discussed by
Farhi, Guth, and Guven \cite{Farhi:1989yr}, who considered tunneling
from a small initial false vacuum seed in asymptotically flat space to
an inflating baby universe inside of a black hole.  They constructed
an instanton for this process, but found that its metric is
degenerate.  The instanton action could still be calculated, but it is
not clear that such pathological instantons are legitimate.  Fischler,
Morgan, and Polchinski \cite{Fischler:1990pk} considered the same
problem using the Hamiltonian formalism and found no inconsistencies.
The nucleation rate they found agrees with the result of
Ref.~\cite{Farhi:1989yr} based on the degenerate instanton.  But this
issue remains controversial.

Nucleation of high-energy inflating regions can also be pictured as a
two-step process.  First a bubble of high-energy vacuum $i$ having
radius $R<H_i^{-1}$ spontaneously nucleates in the parent vacuum $j$,
and then this bubble tunnels to an inflating baby universe contained
inside of a black hole by the process discussed in
Refs.~\cite{Farhi:1989yr,Fischler:1990pk}.  One expects that the rate
for the first step is $\Gamma \sim \exp(-2\pi M/H_j)$, where $M$ is
the mass of the bubble, and the tunneling action is estimated as
\cite{Farhi:1989yr,Fischler:1990pk} $S\sim (\mpl/H_i)^2$.  Farhi et
al.~\cite{Farhi:1989yr} find that the minimal bubble mass required for
the tunneling is $\mpl^2/2H_i$; then the nucleation
  rate is $\Gamma\sim \exp(-\pi \mpl^2/H_i H_j)$.  For $H_i\gg H_j$
  this is the dominant factor determining the nucleation rate of baby
  universes.  This is in agreement with the estimate in
  Eq.~(\ref{flyover}).

Another possible objection to nucleation of inflating baby universes is that it is in conflict with the detailed balance condition (\ref{balance}).  This condition however does not follow from any fundamental principle.  It is violated in particular by transitions between de Sitter and anti-de Sitter vacua, which are necessarily present in any multiverse theory.

As we argued above, inflating baby universes should nucleate at some nonzero rate, even in the absence of instantons, because this process is allowed by all conservation laws.  A calculation of their nucleation rate was attempted in Refs.~\cite{Brown:2011ry,Blanco-Pillado:2019xny}.  This calculation appears to be reliable when the energies of the two vacua and the height of the barrier separating them are all sub-Planckian and are comparable to one another.  But in the opposite limit, when $H_i\gg H_j$, the initial fluctuation is strongly influenced by gravitational effects and calculation of its probability requires a quantum theory of gravity.  Here we shall assume that the nucleation rate in this case is given by Eq.~(\ref{flyover}), which seems to be a plausible guess.

\subsection{Boltzmann brains}

Finally, we expect Boltzmann brains to appear at the rate given by
Eq.~(\ref{eqn:thermal}) with brain mass $\mbb$.  This process is
dominated by the lightest brains that need to be considered for
anthropic reasoning.  These may not actually be brains, per se, but
tiny computers that stimulate human thought sufficiently well to be
considered in anthropics.\footnote{See Ref.~\cite{Schneider:2013iba}
  for some discussion of the difficulty in determining what systems
  should be included in anthropic reasoning.}  We will assume here
  that $\mbb > \mpl \approx 2\times 10^{-5}$ g.  This is correct
  for a human brain and for any computer that we have built so far.
  It is not correct if the only limit is the number of bits that the
  computer can store \cite{DeSimone:2008if}, i.e., if we are not
  concerned with what this computer might be made of and how it can
  operate.  The minimum mass of a working computer is uncertain.  See
  Ref.~\cite{DeSimone:2008if} for further discussion.

\section{The Boltzmann brain problem}\label{sec:BB}

To avoid domination by Boltzmann brains requires that the rate of
Boltzmann brain production $\gbb_i$ is less than the vacuum decay rate
$\Gamma_i = \sum_j \Gamma_{ij}$ in every vacuum $i$
\cite{Bousso:2008hz,DeSimone:2008if}.  Let us review the basic
argument.  Consider some vacuum $i$ in which there are ordinary
observers.  First we rewrite Eq.~(\ref{eqn:rate}),
\be
\frac{df_i}{d\eta} = M_{ij} f_j\,,
\ee
where $M_{ij} =\kappa_{ij} - \delta_{ij} \kappa_i$.  In the limit
where the cutoff grows without bound, this situation can be analyzed
by finding the least negative eigenvalue $-q$ of the matrix $M$ and
the corresponding eigenvector $s$ so that $\sum_j\kappa_{ij}s_j -
\kappa_i s_i = -q s_i$.  The fraction of volume near the cutoff
surface in each vacuum $i$ is then given by $s_i$.  The number of
Boltzmann brains in that vacuum is proportional to $s_i\gbb_i$,
because most of the volume is near the cutoff surface.  Meanwhile, the
number of ordinary observers is proportional to the rate at which new
vacuum of type $i$ is created, which is $\sum_j\kappa_{ij} s_j =
(\kappa_i -q) s_i$.  Ordinary observers are generally found in
collapsed regions, which require some adjustment to the scale factor
measure.  However this adjustment is insignificant compared to the
double-exponential nature of $\Gamma_i$ and $\gbb_i$, so it will not
be important here.

Now $q$ is generally smaller than the total decay rate of the dominant
vacuum, which is less than that of vacuum $i$.  (Since there's only
one dominant vacuum, it's very unlikely that it is able to support
Boltzmann brains.  If it is, Boltzmann brains would certainly dominate
\cite{Bousso:2008hz,DeSimone:2008if}.)  Both $\kappa_i$ and $q$ are tiny numbers, and
generally they are quite far apart.  So $q$ can be ignored and we find
$\kappa_{ij} s_j \approx \kappa_i s_i$, i.e., the rate of creation and
the rate of decay are nearly equal.  We are not concerned with
differences in prefactors, so $\kappa_i$ and $\Gamma_i$ are
interchangeable, and the condition to avoid Boltzmann brain domination
in vacuum $i$ is that $\gbb_i <\Gamma_i$.  Refs.~\cite{Bousso:2008hz,DeSimone:2008if}
show that the condition to avoid Boltzmann brain domination overall
is that $\gbb_i <\Gamma_i$ in every vacuum.

Included in $\kappa_i$ is $\kappa_{0i}$, the rate of formation of
black holes.  The rate for black holes of mass $M$ is proportional to
$\exp(-M/T)$, so it is dominated by the smallest black hole possible.
Let us say this has mass $\mbh$, which is around the Planck mass.
Thus $\kappa_i$ is at least of order $\exp(-\mpl/T)$.  Meanwhile,
$\gbb_i$ is of order $\exp(-\mbb/T)$, where $\mbb$ is the minimum
Boltzmann brain mass.  With our assumption that $\mbb>\mpl$, it
follows that $\gbb_i \ll \Gamma_i$, and there is no problem with
Boltzmann brains.

The question of Boltzmann brain dominance involves comparing the
number of Boltzmann brains and ordinary observers before the cutoff,
so it may be counterintuitive that it is affected by the production of
black holes, which are neither of these.  Here is a way to understand
how this happens.  Consider a multiverse up until a scale factor
cutoff.  Most of the volume is near the cutoff, so we only need to
look there.  Most ordinary observers are in regions that were created
not long before the cutoff and thus still have conditions where
observers can live.  But Boltzmann brains are in regions that were
formed long ago, so we need to know how much volume these regions
have.

Let us put a large but finite number of evenly spaced fiducial
particles on the initial surface, traveling along the geodesics that
we used to define the scale factor measure. In expanding regions,
equal scale factor time means an equal amount of expansion, so each
particle represents the same amount of spatial volume.  Then the ratio
of different volumes is just the relative number of particles that
they contain.

The effect of the black holes is to swallow up some of these particles
so that they do not reach the cutoff surface.  The result is that the
volume of a given vacuum on the cutoff surface is smaller than it would
be without black hole formation.  Thus $s_i$, the fraction of the cutoff
surface in volume $i$, is inversely proportional to the decay rate
$\Gamma_i$.  Black hole nucleation increases $\Gamma_i$ and so decreases
$s_i$.  The number of Boltzmann brains in this vacuum is then
proportional to $\gbb/\Gamma_i$.  This leads to the criterion used
above.

The process of removing particles by black hole formation is extremely
slow.  A Planck-scale black hole removes only a fraction of order
$H^3/\mpl^3$ of the Hubble volume where it forms. More importantly,
such a black hole only occurs once in every $\exp(\mbh/T)$ Hubble
volumes.  Thus we must wait time of order $\exp(\mbh/T)$ Hubble times
before this effect is important.  During this time the universe
expands by a factor $\exp(\exp(\mbh/T))$.  In our present universe,
this is about $\exp(\exp(10^{60}))$, a remarkably large number.
Nevertheless, the scale factor measure instructs us to consider the
limit where the scale factor goes to infinity, so the required scale
factor to reach a steady-state situation does not matter.

\section{Vacuum dynamics}\label{sec:vacuum}

The possibility of less-suppressed tunneling to higher energy vacua
changes the distribution of different possible states and thus the
results to be expected under anthropic reasoning.  The fraction of the
volume in some vacuum $i$ according to the scale factor measure
depends on the tunneling rates to get from the dominant vacuum to
vacuum $i$ \cite{SchwartzPerlov:2006hi}.  To reach any anthropically
allowed vacuum from the dominant vacuum we generally need an upward
jump, or many such jumps, followed by many downward
jumps.\footnote{The dominant vacuum is likely to have a very low
  supersymmetry breaking scale $\eta_*$.  Its energy density
  $U_*\lesssim \eta_*^4$ is then likely to be extremely small.  It is
  also reasonable to expect that this nearly supersymmetric vacuum can
  support neither ordinary observers nor Boltzmann brains.  For a
  discussion of the expected properties of the dominant vacuum in
  string theory see Ref.~\cite{Douglas:2012bu} and references therein.}
Which vacua are easily reached depends on which process we consider.

If we consider only the Lee-Weinberg process, there is a large
suppression factor given by Eq.~(\ref{balance}).  This suppression is
less important when the two vacua are close in energy.  Thus the
favored vacua are those which can be reached from the dominant vacuum
by small upward jumps followed by downward jumps.  Depending on the
structure of the landscape, these vacua may be sparse enough that the
anthropic explanation of the cosmological constant does not work
\cite{SchwartzPerlov:2006hi,Olum:2007yk}.

However, when we consider the formation of small regions of
high-energy vacuum $j$, the mass of the region is $\mnuc_j \sim
\mpl^2/H_j$ and the suppression goes as $\exp(-\mpl^2/(H_i H_j))$.
Thus the least suppressed transitions are those to the largest
$H_i$.  Furthermore, there is little dependence on which is the
dominant vacuum, because wherever one starts, the same high-energy
vacua are preferred.  From those vacua, one must then drop, generally
in a number of steps, to the anthropic region.  This pattern
of transitions generally leads to a much smoother distribution
of probabilities for the different vacua \cite{Clifton:2007bn}.

\section{Conclusion}\label{sec:conclusions}

In an eternally inflating universe, there is the possibility of
Boltzmann brain domination, meaning that anthropic reasoning would
lead to the nonsensical conclusion that we are Boltzmann brains.  In
the scale factor measure, this disaster is avoided when the rate of
Boltzmann brain nucleation is smaller than the vacuum decay rate in
each vacuum (and the dominant vacuum does not support Boltzmann
brains).  If one considers decay only by the Coleman-De Luccia and
Lee-Weinberg processes, this may not be the case (but see
Ref.~\cite{Freivogel:2008wm} for a claim that string theory vacuum
decay rates in string theory are always larger than $\gbb$.).  However
we showed above that black hole nucleation should be included in the
vacuum decay rate, and this process is much less suppressed than
Boltzmann brain production, under a rather mild assumption that the
mass of a Boltzmann brain should be greater than the Planck mass.  Thus
we should not expect to be Boltzmann brains.

We also discussed the nucleation of small regions of inflating
high-energy vacuum.  If vacua of high enough energies exist, this
process also would prevent Boltzmann brain domination.  In any case it
modifies the probability distribution of the various vacua, likely
giving a more uniform distribution for different anthropic
possibilities and guaranteeing that anthropic explanations of the
smallness of the cosmological constant are not affected by highly
nonuniform probability distributions across anthropic vacua.

We finally mention the swampland conjectures which have been
intensively discussed in recent years (see Ref.~\cite{Bedroya:2020rac} for
an up-to-date review and references).  According to these conjectures
metastable de Sitter vacua do not exist and many models of eternal
inflation are also ruled out.  However, it was shown in
Ref.~\cite{Blanco-Pillado:2019tdf} that eternal inflation driven by
inflating domain walls may still be possible.  It would be interesting
to apply the considerations of the present paper to this kind of
multiverse models.

\section*{Acknowledgments}

We are grateful to Jose Juan Blanco-Pillado, Heling Deng, Michael
Douglas, and Ben Freivogel for useful discussions.  This work was
supported in part by the National Science Foundation under grant
No. 1820872.

\maketitle

\bibliography{paper}

\end{document}